
\input harvmac
%
%
%
%
%

\parindent=0pt
\Title{SHEP 95-12}{\vbox{\centerline{Two Phases for Compact U(1) Pure Gauge
Theory}
 \vskip4pt\centerline{in Three Dimensions}}}

\centerline{\bf Tim R. Morris}
\vskip .12in plus .02in
\centerline{\it Physics Department}
\centerline{\it University of Southampton}
\centerline{\it Southampton, SO17 1BJ, UK}
\vskip .7in plus .35in

\centerline{\bf Abstract}
\smallskip
We show that if actions more general than the usual simple plaquette action
($\sim F_{\mu\nu}^2$) are considered, then
 compact $U(1)$ {\sl pure} gauge theory in three Euclidean
dimensions can have two phases.
Both phases are confining phases, however in one phase the monopole
condensate spontaneously `magnetizes'. For a certain range of parameters
the phase transition is continuous, allowing the definition of a strong
coupling continuum limit.
We  note that these observations have
 relevance to the `fictitious' gauge field
theories of strongly correlated electron systems, such as those describing
high-$T_c$ superconductors.
\vskip -1.5cm
\Date{\vbox{
{hep-th/9505003}
\vskip2pt{April, 1995.}
}
}
\catcode`@=11 
\def\slash#1{\mathord{\mathpalette\c@ncel#1}}
 \def\c@ncel#1#2{\ooalign{$\hfil#1\mkern1mu/\hfil$\crcr$#1#2$}}
\def\lsim{\mathrel{\mathpalette\@versim<}}
\def\gsim{\mathrel{\mathpalette\@versim>}}
 \def\@versim#1#2{\lower0.2ex\vbox{\baselineskip\z@skip\lineskip\z@skip
       \lineskiplimit\z@\ialign{$\m@th#1\hfil##$\crcr#2\crcr\sim\crcr}}}
\catcode`@=12 

\def\phi{\varphi}

\def\epsilon{\varepsilon}
\def\bi{{\bf i}}
\def\j{{\bf j}}
\def\p{{\bf p}}

\def\x{{\bf x}}
\def\y{{\bf y}}

\def\B{{\bf B}}

\def\J{{\bf J}}
\def\div{\nabla.}
\def\D{{\cal D}}
\def\L{{\cal L}}
\def\Z{{\cal Z}}

\def\frac#1#2{{#1\over#2}}
\def\Fe{F_{\mu\nu}}

\def\rline{{\rm I}\!{\rm R}} 
\parindent=15pt

As shown by Polyakov\ref\polyi{A.M. Polyakov, Phys. Lett. 59B
(1975) 82.}\ref\polyii{A.M. Polyakov, Nucl. Phys. B120 (1977) 429.},
there is a world of difference between {\sl compact}
and {\sl non-compact} $U(1)$ pure gauge theory: in the non-compact
case the $U(1)$
gauge transformations (and correspondingly the
bare connections $A_\mu$) are valued on the whole real line,
 while in the compact case they
are valued on a circle,\foot{We have in mind a
lattice formulation.}\  
and thus allow
 `magnetic' monopole configurations. These cause three dimensional
compact $U(1)$ pure gauge theory to undergo monopole condensation, resulting
in a
confined disordered phase for all non-zero lattice spacing\ref\mack{This was
proved for the Villain action in -- M. G\"opfert and G. Mack, Commun. Math.
Phys. 82 (1982) 545.}.
However, effectively only the lowest order kinetic term was considered,
corresponding in the
na\"\i ve continuum limit to  $\sim\Fe^2$. In this
letter we consider more general Lagrangians for the compact case,
forming a complement to the study of three dimensional non-compact
pure gauge theory reported
in ref.\ref\ui{T.R. Morris, Southampton preprint SHEP 95-07,
hep-th/9503225.}. Indeed in contrast to that case, we find that
new continuum limits are reachable with more general Lagrangians.

These continuum limits are formed at the phase transition between the
confined disordered phase described above, and an ordered phase in which
the monopole condensate spontaneously `magnetizes'. The
magnetized state appears in the regime where the lowest order kinetic
term  has the `wrong' sign, leading to vacuum instability in the
monopole condensate. Possibly the most interesting physical application
of these ideas are to   recent theories of
 strongly correlated
electron systems, such as those describing  high
temperature superconductors\ref\tc{G. Baskaran and P.W. Anderson, Phys. Rev.
B37
  (1988) 580\semi
P.B. Wiegmann, Phys. Rev. Lett. 60 (1988) 821\semi
G. Baskaran, Physica Scripta. T27 (1989) 53\semi
P.A. Lee, Phys. Rev. Lett. 63 (1989) 680
}--\nref\phases{I. Affleck and J.B. Marston, Phys. Rev. B37 (1988) 3744\semi
G. Kotliar, Phys. Rev. B37 (1988) 3664.}\ref\ioffe{L.B. Ioffe and A.I. Larkin,
Phys. Rev. B39 (1989) 8988.}:
dynamically generated strongly coupled compact $U(1)$
gauge fields naturally arise in their description of
the  effectively planar state in these materials.
At the microscopic level,
an order parameter $\Delta_{\bi\j}=\vev{c^\dagger_{\bi\alpha} c_{\j\alpha}}$,
where $c^\dagger_{\bi\alpha}$ is an electron creation operator of spin $\alpha$
at site $\bi$, plays a central r\^ole; the compact gauge field arises
as the phase $\phi_{\bi\j}$ of this `link field'.
Since these `fictitious' $U(1)$ gauge fields
are born at the microscopic level without kinetic terms, but instead
receive their dynamics through fermionic (i.e. electronic) fluctuations,
the lowest order kinetic term can naturally arise with the
wrong sign. In the simplest case of just nearest neighbour interactions
(and concentrating on the dielectric state with strong on-site repulsion),
mean field approximations indicate that three different phases could
exist:
a `uniform phase' in which the phases may be chosen
so that $\Delta_{\bi\j}=const.$, a `molecular crystal phase' in
which $\Delta_{\bi\j}\ne0$ for only one bond per site, and a `flux phase'
in which $|\Delta_{\bi\j}|=const.$ but the sum of the phases around an
elementary plaquette (called $F^0_{\mu\nu}(\x)$ below) equals
$\pi$ \phases\ioffe.\foot{These
same approximations generally disfavour the flux phase, but
the approximations are not expected to be
reliable for determining the energetics\ioffe.}\
The present work can be regarded as furnishing
a phenomenological Landau Ginzburg
description close to the
flux--uniform phase transition, which goes beyond the mean
field analysis, but in an unrealistic isotropic setting
in which also $|\Delta_{\bi\j}|$ is held fixed and all other quasiparticle
excitations are neglected.

In this respect, we note that
the gauge invariance of the (low energy) fluctuations ensures that
the effective action for the fictitious gauge field is gauge invariant
along the Euclidean
 `time' direction\foot{compactified with circumference inversely
proportional to the temperature}\ also, while
the compactness of the $U(1)$ gauge group guarantees that monopole
configurations (which are instantons of the planar state)
are {\it a priori} allowed. The flux phase precisely corresponds
to maximum magnetization of the monopole condensate along the time
direction, thus the fate of the flux phase and of the monopole gas
are intimately linked.
 The other quasiparticles have a profound effect on
the dynamics of the monopole gas, so that the resulting physics of these
instantons is
not yet clear\ioffe\ref\inq{S. Khlebnikov, Phys. Rev. B50 (1994) 6954\semi
S. Chakravarty, R.E. Norton and O.F. Sylju\aa sen, Phys. Rev. Lett. 74
(1995) 1423.}. The present
formulation may help to clarify the situation, if it can be
 generalised to include the interactions
with the other quasi-particles.

The natural order parameter turns out to be
the `magnetic' field $B_\mu(\x)\equiv{1\over2}
\epsilon_{\mu\nu\lambda} F_{\nu\lambda}(\x)$.
In a cubic lattice regularization, the bare magnetic field corresponds
to the plaquette angle
$$\epsilon_{\mu\nu\lambda}B^0_\lambda(\x)\equiv
F^0_{\mu\nu}(\x)=A^0_\mu(\bi a)+A^0_\nu(\bi a+{\hat\mu}a)
-A^0_\mu(\bi a+{\hat\nu}a)
-A^0_\nu(\bi a)\quad,$$
where $\phi_{\bi,\bi+{\hat\mu}}\equiv A^0_\mu(\bi a)$, $a$ is the lattice
spacing,
$\x=\bi a+{a\over2}({\hat\mu}+{\hat\nu})$
is centred in the elementary plaquette,
and ${\hat\mu},{\hat\nu}$ are unit vectors in the directions $\mu,\nu$.
Reality, gauge invariance and periodicity ensure that any physically sensible
bare action
may be written as a bounded single valued function of the plaquette variables:
$\cos\!\left(B^0_\mu\right)$ 
and
$\ \sin\!\left(B^0_\mu\right)$. 

In gauge theory,
it is usual to think of the partition function $\Z$ as defined by
a functional integral over the gauge field $A_\mu(\x)$. We
take a step backwards however,
and define $\Z$ as a functional integral over
$B_\mu(\x)$, together with the constraint $\partial_\mu
B_\mu=0$ inserted as a functional delta function in the path integral:
\eqn\noncom{\Z^0_{(non-compact)}=\int\!\D \B\,\delta[\,\div\B\,]\,
\e{-{\textstyle{1\over g_0^2}} S_0[\B]}\quad.}
Note that the Jacobian for the change of variables is just a constant
(in an Abelian gauge theory).
The action $S_0$ will be left general for the moment, except that we will
use the fact, mentioned above, that the microscopic
Lagrangian densities are bounded.
We will take them to be normalized so that this bound is of order one;
$g_0$ will thus be analogous to the electromagnetic coupling constant, being
small in the usual Gaussian continuum limit. (We will
assign the natural geometrical
(inverse length) dimension to
$A_\mu$, namely $[A]=1$,  and thus $[B]=2$,
so that in three dimensions $g_0$ has
dimension $[g_0]=\half$).

However in the compact case, we must take account
of apparently singular instanton configurations, corresponding to
monopoles sitting at positions
$\x_s$ with integer
charges $q_s$, which appear
as a result of the fact that the phase of the link field (the bare gauge field)
 is identified under changes of $2\pi$.
We have a choice: we can either keep track of the Dirac
strings, explicitly recalling that these are invisible
to the microscopic Lagrangian when necessary\polyi\polyii,
or we can remove them by using the Wu-Yang prescription\ref\wuy{T.T. Wu
and C.N. Yang, Phys. Rev. D12 (1975) 3845.}, in which case the gauge
field may be chosen to be smooth in patches,
 and identified across the patches by gauge
transformations with non-zero winding number,
that is
$A_\mu(\x)$ is regarded as a connection on a non-trivial $U(1)$
bundle over $\rline^3-\{\x_1,\cdots,\x_N\}$. The two representations
are physically equivalent but we will assume
the Wu-Yang formalism,
because it is more convenient for the continuum limit, and also
emphasises that the quantization of monopole charge, even in this
pure gauge case, is not particular to
the lattice. A DeGrand-Toussaint\ref\degr{T.A. DeGrand and D. Toussaint,
Phys. Rev. D22 (1980) 2478.}\ map to the `physical' bare
magnetic field, by adding integer multiples of $2\pi$ to ensure $-\pi<
B^0_\mu(\x)\le\pi$,  may be regarded as a
lattice Wu-Yang prescription, justifying the statement that
the two view-points are equivalent. We will be implicitly assuming that such a
map has been performed at the lattice level.

We are now ready to consider the changes monopole fluctuations make to
the partition function \noncom.
The natural regime to consider, for example in the theories of high-$T_c$
superconductivity, is $g_0\sqrt{a}\sim1$.
In this regime there is a {\sl freely fluctuating} monopole density of order
unity monopoles per elementary cube (i.e. per volume $a^3$),
so that the bare monopole charge
density $\rho_0(\x)\sim 2\pi \sum_s q_s \delta(\x-\x_s)$ can be assumed to have
a continuum limit $\rho(\x)$. In this case the partition function is simply
given by
\eqn\comuni{\Z=\int\!\D\rho\int\!\D\B\,\,\delta[\,\div\B-\rho\,]\,
\e{-{\textstyle{1\over g^2}} S[\B]}\quad,}
which of course we may integrate to give
\eqn\comunii{\Z=\int\!\D\B\,\,\e{-{\textstyle{1\over g^2}} S[\B]}\quad.}
We will show later that this  partition function is also obtained from the
appropriate limit of the dilute instanton gas approximation\polyi\polyii.
Further justification can be obtained by considering the simplest (Gaussian)
action
\eqn\gau{S=S_{Gaussian}={1\over2}\int\!d^3\!x\, B^2\quad.}
Integrating out the $\B$ field in \comuni\ by writing
$\B\to\B-\nabla \phi$, with $\nabla^2\phi=-\rho$, gives
$$\Z=\Z_{(Gauss)}\,\int\!\D\rho\,\,\exp\left\{-{1\over 8\pi
g^2}\!\int\!d^3\!x\!\int\!d^3\!y\,\,{\rho(\x)\rho(\y)\over |\x-\y|}\right\}
\quad,$$
where $\Z_{(Gauss)}$ is the Gaussian integral over transverse $\B$ (the
photons), analogous to \noncom.
This is nothing but the required continuum limit of the Banks-Myerson-Kogut
formulation of the lattice monopole gas\ref\bmk{T. Banks, R. Myerson and
J. Kogut, Nucl. Phys. B129 (1977) 493.}.
If we substitute \gau\ in \comunii, we see that the disordering effect of
the monopole condensate has been total: there is no propagation.
We have $\vev{B_\mu B_\nu}(\p)=g^2\delta_{\mu\nu}$, which should be compared
to the dilute monopole plasma result\polyii:
\eqn\deb{\vev{B_\mu B_\nu}(\p)=g^2\left(\delta_{\mu\nu}-{p_\mu p_\nu\over
p^2+m^2}\right)\quad,}
confirming that in this case the Debye
correlation length $\xi_d=1/m\sim a$.
The Wilson loop expectation value\ref\wil{K. Wilson, Phys. Rev. D10 (1974)
2445.}:
\eqn\willoo{\exp\{-W[C]\}=\left\langle\exp\left\{i\int_{{\cal A}}\B.d{\bf s}
\right\}\right\rangle\quad,}
where ${\cal A}$ is the minimal area spanning some macroscopic loop, is
easily seen (by completing the square)
to be $W[C]\sim {\cal A}g^2/a$, so that the theory is
confining over distances of order the lattice spacing. These are
the results that are expected in this regime (e.g. from a strong coupling
expansion), as we will further confirm later.

We see that, even if we consider an action which is a general function of
the field strength\ui\ $S=\int\!d^3\!x\, V(B^2)$ we still have no
dynamics. In this case
it is natural to consider a more general action which reintroduces
propagation through higher order derivative terms. In the gauge theories of
high-$T_c$ superconductivity, such further terms can in any case
be expected to be important. Evidently the partition function \comunii,
yields equivalent physics to that of the Heisenberg ferromagnet (viz. $O(3)$
invariant $n$-vector model), and, close to a phase transition
 a sufficiently general effective Landau Ginzburg
description arises from an action of the form
\eqn\gen{S=\int\!d^3\!x\, \left\{{\kappa\over2} B^2
+{1\over2M^2} (\partial_\mu B_\nu)^2
+\lambda^{(4)} B^4+\lambda^{(6)} B^6\right\}\quad.}
A microscopic Lagrangian for which this is the appropriate description is
for example:
$$\eqalign{\L_0(\x)=-\kappa_0{\textstyle\sum_\mu}\cos[B^0_\mu(\x)]
&-{1\over M^2_0}{\textstyle\sum_{\mu,\nu}}\sin[B^0_\mu(\x)]
\sin[B^0_\mu(\x+a{\hat\nu})]\cr &+\lambda^{(4)}_0
\left({\textstyle\sum_\mu}\cos[B^0_\mu(\x)]\right)^2
+\lambda^{(6)}_0 \left({\textstyle\sum_\mu}\cos[B^0_\mu(\x)]\right)^3\quad.}$$
(Here the bare parameters are assumed to be of order unity as explained
previously -- thus for example we can be sure that the monopole gas is always
in the condensed phase since the action (more strictly fugacity)
for a single monopole is of order
unity. In principle terms containing $\partial_\mu B_\mu$ could appear in \gen\
even though microscopically they are forbidden, but these terms correspond
to furnishing an action for the monopole charge density in \comuni\ and thus
to moving away from this deeply confining regime, as we will see later.)
Therefore there are two phases.\foot{Other phase transitions
 are of course
possible (in principle) but would yield only the cubic rotation
group (or subgroup thereof) in the continuum limit.}\
Both phases have a confining monopole condensate, but in one phase the
monopole condensate spontaneously magnetizes and $\vev{B_\mu(\x)}\ne0$.
Physically, it is easy to give
a  picture of what happens microscopically: For $\kappa$ sufficiently
negative (to overcome quantum fluctuations that renormalize $\kappa$
to more positive values) the `energy' (viz. action)
of the monopole changes sign so that
it becomes favourable to produce monopoles from the vacuum. Simultaneously
however, the `force' between monopoles changes sign so that opposite sign
monopoles are actually repelled from each other -- polarizing the vacuum.
This runaway instability continues until it is balanced by the positive
$\lambda$ interactions (or ultimately by the periodicity of the Lagrangian).
At the microscopic level, the Dirac strings
significantly reorder the magnetic field,  so that different
(but physically equivalent) prescriptions
for identifying the monopole charges can give very different qualitative
pictures of the resulting stable state.
The advantage of the version of the DeGrand-Toussaint prescription
we have adopted is that it unties these effects and allows a description
in terms of the smooth order parameter $B_\mu(\x)$.

{}From \gen\ we conclude that, deep in the  confining regime,
for a certain range of parameters the (zero temperature) phase
transition is continuous in the universality class of the three dimensional
$O(3)$ vector
model Wilson fixed point\ref\kogwil{K.G. Wilson and J.Kogut,
Phys. Rep. 12C (1974) 75.}. Outside this range
the transition is first order, and at the boundary we have a tricritical
point with mean-field critical exponents.
Along the continuous phase transition we can define a continuum limit
whose Minkowskian continuation corresponds to a non-unitary
theory of pseudo-vector
glue-balls  (or rather photon-balls) where the $U(1)$ glue is bound
with binding energy of order the cutoff.

 The situation becomes
 more interesting, if we now reduce the coupling constant $g_0$,
moving away from the deeply confining regime.
We will see that the physics smoothly changes into that of the dilute
monopole gas phase, which we now consider.
For $g\sqrt{a}<\!\!<1$,
the semiclassical limit of
the dilute instanton  gas (about some global minimum field $\vev{\B}$)
is a good approximation. In the broken phase
we  shift $\B\mapsto \vev{\B} +\B$, where the
vacuum expectation value is taken to be independent of $\x$.
The {\sl continuum} integration over magnetic
fields with monopole singularities may now be written: 
$$\Z=\sum_{N=0}^\infty {1\over N!}\sum_{\{q_s=\pm1\}}\left(\prod_{s=1}^N\int\!
{d^3\!x_s\over a^3}\right)\zeta^N\int\!\D \B\,\,\delta\!\Big[\div\B-2\pi
{\textstyle\sum_s} q_s\delta(\x-\x_s)\Big]\,\e{-
{\textstyle{1\over g^2}}S[\B]}\quad.$$
Here the fugacity $\zeta$ is given to good approximation by
$\zeta=\e{-\epsilon_0/(g^2 a)}$, where $\epsilon_0/a$ is the
action of one instanton, and  $\epsilon_0$ is a number of order one
which depends on the
couplings in $S$ and the lattice type (and in the broken phase on $\vev{\B}$).
This follows by dimensions and the bounds mentioned earlier.
We have also restricted the monopole charge to $\pm1$ since the fugacity for
higher charges ($\sim\zeta^{q^2}$) is negligable in this limit. (It is worth
remarking that
the physical monopole charge per unit cell, is bounded
 by a lattice-type dependent number -- which is $|q_s|\le2$ for
a cubic lattice\degr).

Expressing the functional delta-function as a
functional Fourier transform, using an auxiliary field $\chi(\x)$, we
can perform the sums above and obtain
$$\Z[J]=\int\!\D(\B,\chi)\,\,\exp\left\{-{1\over g^2}S[\B]+\int\!d^3\!x\,\left[
i\chi\div\B+{2\zeta\over a^3}\cos(2\pi\chi)+\J.\B\right]
\right\}$$
(up to proportionality constants on $\Z$ which we always ignore).
Here we have also introduced a source $\J(\x)$ for $\B$.
Now it is helpful to split $S$ into the bilinear kinetic term
$\half\int\!d^3\!p\, B_\mu(-\p) \Delta^{-1}(p)_{\mu\nu} B_\nu(\p)$
and interactions
$S_{int}[\B]$ (which are order $B^3$ or higher). Shifting $B_\mu\mapsto
B_\mu-ig^2\Delta_{\mu\nu}.\partial_\nu\chi$ we obtain
\eqn\wee{\eqalign{\Z[J]=&\int\!\D(B_\mu,\chi)\,\,\exp\Big\{
-{1\over 2g^2} B_\mu.\Delta^{-1}_{\phantom{-1}\mu\nu}.B_\nu
-{1\over g^2}S_{int}[B_\mu-ig^2\Delta_{\mu\nu}.\partial_\nu\chi]\cr
&-{g^2\over2}\int\!d^3\!x\,
\left[\partial_\mu\chi\,\Delta_{\mu\nu}.\partial_\nu\chi-
{m^2\over 4\pi^2}\cos(2\pi\chi)\right]
+J_\mu.\left(B_\mu-ig^2\Delta_{\mu\nu}.\partial_\nu\chi\right)\Big\}\quad,}}
where we have introduced the Debye mass $m^2=8\pi^2\zeta/(g^2a^3)$. Tracing
the factors of $g$,  one can see that in this form the theory is manifestly
weakly coupled. (For this it is helpful to note that $m^2/g^4$ is exponentially
small in this regime).

If we specialize to the Gaussian action \gau, then \wee\ neatly
 summarises Polyakov's solution. To see this, note that in this case
$\Delta_{\mu\nu}=\delta_{\mu\nu}$
and $S_{int}=0$; $\chi$ is (up to a factor $2\pi$) the Debye-Huckel potential
field used in ref.\polyii. The equivalence is completely clear if we write
$\J=i\nabla\eta+{\tilde\J}$, where $\eta$ is Polyakov's source for monopole
charge density and the transverse photon
 source satisfies $\div{\tilde\J}=0$, and then
integrate out $\B$ to obtain:
$$\eqalign{\Z=\int\!\D\chi\,\,\exp\Bigg\{{g^2\over2}\int\!d^3\!p\,
{\tilde J}_\mu(-\p)
&\left[\delta_{\mu\nu}-{p_\mu p_\nu\over p^2}\right]{\tilde J}_\nu(\p)\cr
&-{g^2\over2}\int\!d^3\!x\,
\left[\left(\partial_\mu[\chi-\eta]\right)^2-
{m^2\over 4\pi^2}\cos(2\pi\chi)\right] \Bigg\}\quad.} $$
It follows  of course that for the Gaussian
action one obtains the same results from \wee\
as obtained in refs.\polyi\polyii,
namely $m$ is indeed the Debye mass as stated above and defined in eq.\deb,
and we have confinement: $W[C]\sim m g^2 {\cal A}$.

Now note that if we put $g\sqrt{a}=const.<\!\!<1$, then the
instanton computation remains valid, but $m\propto 1/a$. At low energies
(equivalent to $a\to0$), this is the deeply
confining and disordered regime
we discussed previously. We see that the large effective
Debye mass ensures that
the contributions from the $\chi$ field are negligable
for the low energy excitations, and the partition function \wee\ reduces to
\comunii. Also the Gaussian results stated above go over to those deduced from
\comunii\ as they should. This provides our final justification for the
effective partition function \comunii.

Now we briefly survey
 the results one obtains for the general actions such as
\gen, away from deep confinement. Firstly, it is not hard to show [by e.g.
changes of variables on the  quadratic parts of \wee] that the effective
susceptibility (propagator)  $\Delta_{eff}$ for the magnetic field,
is  given  generally, to lowest order in $g$, by:
\eqn\deff{\Delta^{-1}_{eff}(p)_{\mu\nu} ={1\over g^2}
\left\{\Delta^{-1}(p)_{\mu\nu}+{p^\mu p^\nu\over m^2}\right\}\quad.}
In the symmetric phase (and for small $g$) we may take $\kappa=1$ in \gen\ by a
redefinition of $g$, so that $\Delta^{-1}_{\phantom{-1}\mu\nu}\equiv
\delta_{\mu\nu}(1+p^2/M^2)$. This yields
$$\Delta_{eff}(p)_{\mu\nu}=\left(\delta_{\mu\nu}-{p_\mu p_\nu\over p^2}\right)
{g^2M^2\over p^2+M^2}+{p^\mu p^\nu\over p^2}{g^2 m^2_{eff}\over p^2+m^2_{eff}}
\quad,$$
where $m_{eff}^2=m^2M^2/(m^2+M^2)$. This reduces to the Debye formula \deb\ in
the limit $M\to\infty$ as it should, however we see that generally the
transverse susceptibility responds according to the mass $M$ of the
`pseudovector glueball'
as expected from \gen, but the longitudinal susceptibility
behaves as a bound state, of the longitudinal parts of the pseudo-vector
excitation and the Debye mass `scalar glueball', with a mass $m_{eff}$
which is always less than $m$ or $M$ (and greater than min$[m,M]/\sqrt{2}$).

In the broken phase we take without loss of generality
$\Delta^{-1}_{\phantom{-1}\mu\nu}\equiv n_\mu n_\nu+p^2\delta_{\mu\nu}/M'^2$,
where ${\bf n}$ is the unit vector in the direction $\vev{\B}$.
Now the susceptibility has three eigen-directions: The transverse magnon
(i.e. along the $\p\times\vev{\B}$ direction) remains massless, but in the
$\vev{\B}$ -- $\p$ plane two new directions are distinguished with
susceptibilities which are no longer simple poles but of the form
$2g^2m'^2_{eff}/S_\pm$, where $S_\pm=m'^2_{eff}+p^2+\sqrt{p^4+2m'^2_{eff}\,
p^2\cos{2\theta}+m'^2_{eff}}$. Here $\theta$ is the
angle between $\p$ and $\vev{\B}$
and the effective mass is the equivalent to that of the unbroken phase:
$m'^2_{eff}=m^2M'^2/(m^2+M'^2)$.

{}From \deff, these leading order changes to the susceptibility can be
incorporated by changing the partition function \comunii\ by
\eqn\effi{S[\B]\mapsto S[\B]+\half\xi_d^2\int\!d^3\!x\,(\div\B)^2\quad,}
but at higher order in $g$ the effective magnetic field action also inherits,
from the $\chi$ dynamics in \wee,
non-local changes (the width of the bound state)
proportional to factors of $\div\B$.

It would be
interesting to understand what effect the change to weak confinement has
on the deeply confining phase transition considered earlier (that is
assuming the parameters are tuned so that $\xi_d$  diverges
with the correlation length). Can it still be continuous, and if so
in what universality class? These questions could be addressed within the
epsilon expansion\kogwil\ starting from \wee, although it is not  clear
that the epsilon expansion should be reliable here. The first corrections
to deep confinement, i.e. where $p/m<\!\!<1$, come from allowing $\div\B$
terms in \gen, the correction to the quadratic part being given by \effi.
[They correspond to furnishing an action for the monopole
charge density in \comuni].
The effect of these corrections on the phase transition could be investigated
by both the epsilon expansion and the derivative expansion\ref\lat{See e.g.
ref. \ui\ and  T.R. Morris, Lattice '94 conference report,  SHEP 94/95-10,
hep-lat/9411053.}. Since we found no continuous phase transition for
general actions in non-compact pure gauge $U(1)$ theory\ui, it must be that
for sufficiently weak confinement the smooth phase transition discussed
earlier, disappears. The simplest assumption is that it becomes first
order. This implies that
 the  non-compact case also has two phases, with
$\B$ being the order parameter, but that the non-compact case phase transition
is always first order. Since mean field theory allows for continuous
phase transitions, this means that a
Coleman-Weinberg mechanism operates here
as in the classic case of scalar QED:
fluctuations drive the
non-compact transition first order\ref\colw{S. Coleman and E. Weinberg,
Phys. Rev. D7 (1973) 1888\semi
B.I. Halperin, T.C. Lubensky and S.K. Ma, Phys. Rev. Lett. 32 (1974) 292.}.
But another possibility is that for $\xi_d$ sufficiently large a new phase
opens up in which the $\B$ field becomes disordered independently of the
effects of the monopole plasma (and
not therefore unravelable by DeGrand-Toussaint transformations).
This possibility was conjectured recently in the context of three dimensional
non-compact QED\ref\dhok{D. Cangemi, E. D'Hoker and G. Dunne,
Phys. Rev. D51 (1995) 2513.}. Presumably
in this phase $\vev{\B}$
would still vanish, and  the relevant order parameter would have to be
 composite e.g. $B^2$. This could be investigated by extending the analysis of
the pure gauge non-compact case\ui\ to allow for such a possibility.

\bigbreak\bigskip\bigskip\centerline{{\bf Acknowledgements}}\nobreak
It is a pleasure to thank
Ian Aitchison, Simon Hands and Mike Teper for helpful discussions,
and the SERC/PPARC for providing financial support through an Advanced
Fellowship.

\listrefs

\end